\documentclass[%
final,
reprint,
superscriptaddress,
showpacs,
amsmath,
amssymb,
aps,
pra,
]{revtex4-1}
    
\usepackage{graphicx}% Include figure files
\usepackage{dcolumn}% Align table columns on decimal point
\usepackage{bm}% bold math    
\usepackage[english]{babel}
    
\usepackage{environ}      
\NewEnviron{eqnalign}{\begin{equation}\begin{aligned}\BODY\end{aligned}\end{equation}}

\vspace{1cm}
         
\usepackage{color}
\usepackage{xcolor}
\usepackage[textsize=small]{todonotes}

\NewEnviron{todomini}[1][2do]{\todo[inline,caption={#1}]{  
\begin{minipage}{\textwidth-4pt}  
\BODY
\end{minipage}}}  

\RequirePackage{mathtools}
\RequirePackage{graphicx} 

\RequirePackage{epstopdf}
\RequirePackage{siunitx}
\RequirePackage[utf8]{inputenc}  
\RequirePackage[T1]{fontenc}
\RequirePackage{xspace}   
\RequirePackage[super]{nth}

\RequirePackage{braket}   

\RequirePackage{ifdraft}
\newcommand\source[1]{\ifoptionfinal{}{\textcolor{green}{#1}}}
\newcommand\sourceft[1]{\ifoptionfinal{}{\footnote{\textcolor{green}{#1}}}}
	\RequirePackage[
	bookmarks=true,
	linkbordercolor={1 1 1},
	urlbordercolor={1 1 1},
	citebordercolor={1 1 1},
	colorlinks=true,
	urlcolor=blue,
	citecolor=blue,
	linkcolor=blue,
	anchorcolor=green,
	breaklinks=true,
	linktocpage=true % if true, page numbers instead of lines are blue in the toc,
	]
	{hyperref}

\ifoptionfinal{}{
\RequirePackage[right]{showlabels}     

}  

\RequirePackage{cleveref} % Combined usage of hyperref and cleveref
\crefname{figure}{Fig.}{Figs.}
\crefname{table}{Tab.}{Tab.}
\crefname{equation}{Eq.}{Eqs.}
\crefname{chapter}{Chapter}{Chapters}
\crefname{section}{Sec.}{Sec.}
\crefname{appendix}{Appendix}{Appendices}
\crefname{appchap}{Appendix}{Appendices}
\crefname{appsec}{Appendix}{Appendices}

 \ifoptionfinal{
 \RequirePackage[
nowarn,
nomain,
acronym,      %generate acronym listing   -> Not used in this example (see line with %%% )
toc          %show listings as entries in table of contents
]{glossaries}      
\glsdisablehyper

 }
 {
\RequirePackage[
nomain,
acronym,      %generate acronym listing   -> Not used in this example (see line with %%% )
toc          %show listings as entries in table of contents
]{glossaries}

}

\DeclareDocumentCommand{\captionmacro}{ O{} O{(Color online)} m
}{\caption[#1]{#2 \textbf{#1} #3}}

\DeclareDocumentCommand{\fm}{ m }{\ifmmode #1 \else $#1$\fi}
\def\upket{\fm{\ket{\uparrow}}\xspace}
\def\downket{\fm{\ket{\downarrow}}\xspace}

\def\auxket{\fm{\ket{\mathrm{aux}}}\xspace}

\def\Mgtf{\fm{^{25}\mathrm{Mg}^{+}}\xspace}
\def\Mgtv{\fm{^{24}\mathrm{Mg}^{+}}\xspace}  
\def\Mghd{\fm{^{24}\mathrm{MgH}^{+}}\xspace}

\DeclareDocumentCommand{\ion}{ O{ } O{+}  
m }{\fm{^\mathrm{#1}\mathrm{#3}^\mathrm{#2}}}

\def\dsoh{\fm{{}^2\mathrm{S}_{1/2}}\xspace}  
\def\dpoh{\fm{{}^2\mathrm{P}_{1/2}}\xspace}
\def\dpth{\fm{{}^2\mathrm{P}_{3/2}}\xspace}

\def\bsfull{\fm{\ket{{}^2\mathrm{S}_{1/2}, F=3,m_F=3}}\xspace} 
\def\dsfull{\fm{\ket{{}^2\mathrm{S}_{1/2}, F=2,m_F=2}}\xspace} 
\def\auxfull{\fm{\ket{{}^2\mathrm{S}_{1/2}, F=3, m_F=2}}\xspace}
\def\nbar{\fm{\bar{n}}\xspace}

\def\nbar{\fm{\bar{n}}\xspace}

\def\mus{\fm{\mu}\second}

\def\MHz{\mega\hertz} 

\def\omegaT{\fm{\omega_\mathrm{T}}\xspace}

\newcommand{\subplotlabel}[1]{(\textbf{#1})}

\DeclareDocumentCommand{\commutator}{ m m }{\ifmmode
\left[#1\mathrm{,}\,#2\right] \else $\left[#1\mathrm{,}\,#2\right]$\fi}
\makeglossaries                                 
\makeatletter
\let\oldmakefirstuc\makefirstuc
\renewcommand*{\makefirstuc}[1]{%
  \def\gls@add@space{}%
  \mfu@capitalisewords#1 \@nil\mfu@endcap
}
\def\mfu@capitalisewords#1 #2\mfu@endcap{%
  \def\mfu@cap@first{#1}%
  \def\mfu@cap@second{#2}%
  \gls@add@space
  \oldmakefirstuc{#1}%
  \def\gls@add@space{ }%
  \ifx\mfu@cap@second\@nnil
    \let\next@mfu@cap\mfu@noop
  \else
    \let\next@mfu@cap\mfu@capitalisewords
  \fi
  \next@mfu@cap#2\mfu@endcap
}
\makeatother      
\newacronym{PPLN}{PPLN}{periodically-poled Lithium Niobate}

\newacronym{mode:ip}{ip}{in-phase}
\newacronym{mode:op}{op}{out-of-phase}

\newacronym{PTB}{PTB}{Physikalisch-Technische Bundesanstalt}

\newacronym{AEC}{AEC}{absorption-emission cycle}

\newacronym{DDS}{DDS}{direct digital synthesizer}
\newacronym{PFC}{PFC}{phase frequency comparator}
\newacronym{PI}{PI}{proportional-integral}
\newacronym{TTL}{TTL}{transistor-transistor logic}

\newacronym{ECDL}{ECDL}{external cavity diode laser}
\newacronym{AOM}{AOM}{acousto-optical modulator}
\newacronym{EOM}{EOM}{electro-optical modulator}

\newacronym{PBS}{PBS}{polarizing beam splitter}
\newacronym{GTP}{GTP}{Glan-Thompson prism}

\newacronym{PD}{PD}{photodiode}
\newacronym{PMT}{PMT}{photon multiplier tube}
\newacronym{Nd:YAG}{Nd:YAG}{neodymium-doped yttrium aluminum garnet}

\newacronym{STIRAP}{STIRAP}{stimulated Raman adiabatic passage}
\newacronym{SBC}{SBC}{sideband cooling}
\newacronym{DC}{DC}{Doppler cooling}
\newacronym{BSB}{BSB}{blue sideband}
\newacronym{RSB}{RSB}{red sideband}
\newacronym{LFMS}{LFMS}{laser fluorescence mass spectroscopy}

\newacronym{PRS}{PRS}{photon recoil spectroscopy}
\newacronym{QLS}{QLS}{quantum logic spectroscopy}
\newacronym{QLT}{QLT}{quantum logic technique}
\newacronym{RWA}{RWA}{rotating wave approximation}
\newacronym{LDA}{LDA}{Lamb-Dicke approximation}
\newacronym{TPS}{TPS}{two-point sampling}

\newacronym{EIT}{EIT}{electromagnetically induced transparency}

\newacronym{FWHM}{FWHM}{full width at half maximum}

\newacronym{LIF}{LIF}{laser induced fluorescence}
\newacronym{LAS}{LAS}{laser absorption spectroscopy}

\newacronym{QPN}{QPN}{quantum projection noise}
\newacronym{EST}{EST}{electron shelving technique}

\newacronym{CDF}{CDF}{cumulative distribution function}
\newacronym{QED}{QED}{quantum electrodynamics}

\newacronym{SNR}{SNR}{signal-to-noise ratio}
\newacronym{RF}{RF}{radio frequency}

\newacronym{BBR}{BBR}{blackbody radiation}

\newacronym{MCWF}{MCWF}{Monte Carlo wave-function}
\newacronym{MC}{MC}{Monte Carlo}
 
\begin{document}
%\graphicspath{{media/}}   

\preprint{APS/123-QED}  
  
\title{Efficient sympathetic motional ground-state cooling of a molecular ion}  

% Author list   
\author{Yong Wan}
\affiliation{QUEST Institut, Physikalisch-Technische Bundesanstalt, 38116  
Braunschweig, Germany}  
\author{Florian Gebert}
\affiliation{QUEST Institut, Physikalisch-Technische Bundesanstalt, 38116  
Braunschweig, Germany}    
\author{Fabian Wolf}
\affiliation{QUEST Institut, Physikalisch-Technische Bundesanstalt, 38116  
Braunschweig, Germany}          
\author{Piet O. Schmidt}\email{piet.schmidt@quantummetrology.de}
\affiliation{QUEST Institut, Physikalisch-Technische Bundesanstalt, 38116  
Braunschweig, Germany}  
\affiliation{Institut f\"ur Quantenoptik, Leibniz Universit\"at Hannover, 30167
Hannover, Germany}

\date{\today}
  
\begin{abstract}
Cold molecular ions are promising candidates in various fields ranging from precision spectroscopy and test of fundamental physics to ultra-cold chemistry. Control of internal and external degrees of freedom is a prerequisite for many of these applications. Motional ground state cooling represents the starting point for quantum logic-assisted internal state preparation, detection, and spectroscopy protocols. Robust and fast cooling is crucial to maximize the fraction of time available for the actual experiment. We optimize the cooling rate of ground state cooling schemes for single \Mgtf ions and sympathetic ground state cooling of \Mghd. In particular, we show that robust cooling is achieved by combining pulsed Raman sideband cooling with continuous quench cooling. Furthermore, we experimentally demonstrate an efficient strategy for ground state cooling outside the Lamb-Dicke regime. %Operating in this regime is unavoidable when the Doppler cooling linewidth is much larger than the motional frequencies or even desirable in systems designed for a high sensitivity to displacement forces.
% \begin{description}    
% Secondary publications and information retrieval purposes.
% \item[PACS numbers]
% May be entered using the \verb+\pacs{#1}+ command.
% You may use the \texttt{description} environment to structure your abstract;
% use the optional argument of the \verb+\item+ command to give the category of each item. 
% \end{description}
\end{abstract}

% \pacs{Valid PACS appear here}% PACS, the Physics and Astronomy
\maketitle 
  
% Main sections      
\section{Introduction}  
\label{sec:motivation_sbc}    
Coherent manipulation of quantum systems has been a goal for a wide range of
fields in physics and chemistry. The ability to control internal and external
degrees of freedom of atoms has enabled applications such as quantum computation
\citep{leibfried_experimental_2003, haffner_quantum_2008,
blatt_entangled_2008, wineland_nobel_2013}, quantum simulation \citep{blatt_quantum_2012, schaetz_focus_2013}, and quantum metrology \citep{schmidt_spectroscopy_2005, roos_designer_2006, rosenband_frequency_2008, hempel_entanglement-enhanced_2013, wan_precision_2014, ludlow_optical_2014}.
The control over the internal degrees of freedom of the atomic species is often accomplished through coherent manipulation using laser, microwave or radiofrequency pulses. The external degrees of freedom are typically controlled with various laser cooling techniques. Extending these coherent manipulation techniques to molecular ions promises novel applications including multi-qubit quantum memories, quantum information processing \cite{mur-petit_toward_2013, shi_microwave_2013}, and ultracold chemistry \citep{willitsch_chemical_2008, willitsch_cold_2008, staanum_probing_2008}. Moreover, molecular ions are particularly promising candidates for several spectroscopic tests of fundamental physics, ranging from a possible variation of the electron-to-proton mass ratio \citep{schiller_tests_2005, flambaum_enhanced_2007, beloy_rotational_2011, kajita_estimated_2011, kajita_test_2014} over an enhanced sensitivity to a possible dipole moment of the electron \citep{ravaine_marked_2005, meyer_candidate_2006, petrov_theoretical_2007, leanhardt_high-resolution_2011, loh_precision_2013} to parity violation observable through a small energy difference in enantiomers of polyatomic molecules \citep{berger_chapter_2004, quack_high-resolution_2008, bast_analysis_2010, darquie_progress_2010}. 

However, controlling the quantum state of molecular ions represents a challenge arising from their rich internal level structure, including rotation and vibration. As a result of the absence of selection rules for vibrational transitions (except for a few special cases with near-diagonal Franck-Condon matrices \citep{nguyen_challenges_2011, nguyen_prospects_2011, lien_broadband_2014}) no closed cycling transitions are available to implement laser cooling. This limitation can be overcome by sympathetic cooling with an atomic species trapped together with the molecule. 
%For ions that are not directly accessible to sideband cooling, such as molecular ions, or where cooling has undesired effects, such as in storage qubits in quantum algorithms \citep{kielpinski_architecture_2002}, sympathetic cooling via a co-trapped cooling ion provides an efficient mechanism for preparing the motional ground state. 
%Owing to the strong mutual Coulomb interaction, the motion of the ions is no longer independent, but rather described in the form of normal modes \citep{?}. 
While sympathetic Doppler cooling of molecular ions has been demonstrated by several groups in the past \cite{baba_cooling_1996, welling_ion/molecule_1998, roth_production_2005, tong_sympathetic_2010, goeders_identifying_2013}, only recently sympathetic ground state cooling of CaH$^+$ using Ca$^+$ has been achieved \citep{rugango_sympathetic_2014}. The approach is identical to previous successful implementations using atomic ions trapped in linear radio-frequency Paul traps \citep{rohde_sympathetic_2001, barrett_sympathetic_2003, schmidt_spectroscopy_2005, chou_frequency_2010, home_memory_2009, lin_sympathetic_2013}.

Besides the lowest achievable mean population of motional states, $\nbar$, which can reach $\nbar\sim 0.001$ for a single ion \citep{roos_quantum_1999}, the time spent on cooling is an important aspect. Short cooling times reduce overhead in the experimental cycle and are relevant for quantum algorithms \citep{kielpinski_architecture_2002, lin_sympathetic_2013}, as well as spectroscopy experiments requiring ground state cooling, such as quantum logic spectroscopy \citep{schmidt_spectroscopy_2005, rosenband_frequency_2008} and photon recoil spectroscopy \citep{wan_precision_2014}.
%, or the non-destructive detection of the internal state of a molecular ion using an state-dependent oscillating dipole force \cite{}.

Cooling to the motional ground state of a single mode can be achieved via \textit{resolved sideband cooling} (SBC), which requires the motional sidebands in the excitation spectrum of the system to be spectrally resolved. It has been demonstrated in various systems ranging from single trapped ions in Paul traps \citep{diedrich_laser_1989, monroe_resolved-sideband_1995, roos_controlling_2000, eschner_laser_2003} to neutral atoms in optical lattices \citep{hamann_resolved-sideband_1998, vuletic_degenerate_1998} or optical tweezers \citep{thompson_coherence_2013} and micromechanical oscillators \citep{schliesser_resolved-sideband_2008, teufel_sideband_2011, chan_laser_2011}. 
In the following we will focus on ground state cooling of trapped and localized one- and two-ion crystals, represented by 2-level systems with two metastable electronic states ($\upket$, $\downket$). The crystals are confined in a harmonic trap with motional Fock states $\ket{n}$, separated by the trap frequency $\omegaT$, as shown in Fig.~\ref{fig:principle_sbc}(a). Kinetic energy is removed from the system by selectively driving red-sideband (RSB) transitions involving a change of the electronic state while reducing the motional quantum number, followed by spontaneous emission back to the electronic ground state. In the Lamb-Dicke regime, 
% the recoil energy $\Erec$ is much smaller than the spacing of harmonic oscillator levels, $\Erec\ll\omegaT$, 
changes in the motional state upon spontaneous emission are strongly suppressed \citep{wineland_laser_1979, javanainen_laser_1981, stenholm_semiclassical_1986}. 

Sideband cooling is typically implemented in two flavours. In continuous sideband cooling, a quench laser coupling the metastable excited state (\upket) to a short-lived state is applied simultaneously with the RSB laser, effectively broadening the linewidth of the excited state to optimize the cooling rate \citep{diedrich_laser_1989, marzoli_laser_1994, roos_quantum_1999}. In pulsed sideband cooling, the RSB and the quench (also called repump) lasers are applied sequentially \citep{heinzen_quantum-limited_1990, monroe_resolved-sideband_1995}. 
%Ideally, RSB transfer pulse should be close to a complete inversion of population between the motional levels $\ket{n}$ and $\ket{n-1}$ for good cooling performance. Since the RSB coupling strength depends on the motional state of the mode to be cooled (see Fig.~\ref{fig:Rabi_freq_mg25}) and possibly other so-called ``spectator modes'' \cite{wineland_experimental_1998, leibfried_quantum_2003}, the pulse time needs to be adjusted for each initial motional state. In a sequence of pulses addressing all motional states of a single mode in descending order, the population is ``swept'' all the way to the ground state, where the interaction with the laser stops. 
%However, outside the Lamb-Dicke regime
%for sufficiently large motional excitation, where the size of the ion's wavepacket becomes comparable to the effective cooling wavelength, 

The RSB coupling strength depends strongly on the motional state \cite{wineland_laser_1979, wineland_experimental_1998} and outside the Lamb-Dicke regime even exhibits points of vanishing coupling for certain initial motional states, effectively disabling further cooling beyond these points (see Fig.~\ref{fig:Rabi_freq_mg25}). While this regime has been studied 
%This regime is reminiscent of the cooling dynamics outside the Lamb-Dicke regime, which has been studied 
theoretically \citep{morigi_ground-state_1997, stevens_simple_1998, morigi_laser_1999}, no experimental investigations are known to us.

Here, we demonstrate fast and robust sympathetic ground state cooling of a \Mghd
molecular ion along one direction of motion using a \Mgtf cooling ion. Starting
with a Doppler-cooled single trapped \Mgtf ion, we investigate in \cref{sec:sbc_of_single_ion} a novel repumping scheme for pulsed Raman sideband cooling, in which the excited electronic state in the cooling cycle is effectively quenched to the electronic ground state via coupling to a short-lived excited state. %By exploring the transition between pulsed and continuous sideband cooling,
We demonstrate that the requirements on meeting the optimum RSB pulse length are significantly relaxed for the new quasi-continuous scheme compared to conventional pulsed Raman SBC. As a consequence of the large linewidth of the Doppler cooling transition in \Mgtf, a significant amount of motional state population is trapped above the point of vanishing coupling strength for RSBs. We employ second order RSB transitions to sweep the population beyond this point and demonstrate in \cref{sec:low_trap_frequency} that ground state cooling with Lamb-Dicke factors as large as 0.45 and motional level populations up to $n\sim 120$ becomes possible by employing RSBs up to \nth{8} order. 
%We find that a pulse length matching the maximum of the RSB excitation rate and an equal amount of time spent on first and second order RSBs provides the fastest cooling (\cref{sec:sbc_of_single_ion}). 
By optimizing the ground state cooling rate in terms of pulse lengths for \nth{1} and \nth{2} order RSB and the time spent on cooling with each sideband order, a total cooling time as short as $500~\mu$s for cooling a single ion from Doppler temperature ($\nbar\sim 10$) to the motional ground state is demonstrated.  The experimental results are supported by numerical Master equation simulations of the system (Appendix~\ref{app:numerical_simulation_SBC}). In \cref{sec:sbc_two_ion_crystal} this scheme is extended to cool a \Mghd/\Mgtf ion crystal to the ground state by interleaved cooling of both axial modes. 
%We observe that cooling the modes for the same amount of time provides the shortest total cooling time of 2.5~ms and find a mean residual motional excitation of 0.06(3) and 0.03(3) for the in- and out-of-phase modes, respectively. 
After optimizing the RSB pulse length for each mode and the time spent on cooling each, a total cooling time of 2.5~ms is achieved, resulting in a mean residual motional excitation of 0.06(3) and 0.03(3) for the in- and out-of-phase mode, respectively.

\section{SBC of a single ion}
%####################################################################################
\label{sec:sbc_of_single_ion}  
%####################################################################################
\subsection{Experimental setup}     
%####################################################################################
\label{sec:setup}  
\begin{figure}  
	\centering
	\includegraphics[width=0.6\columnwidth]{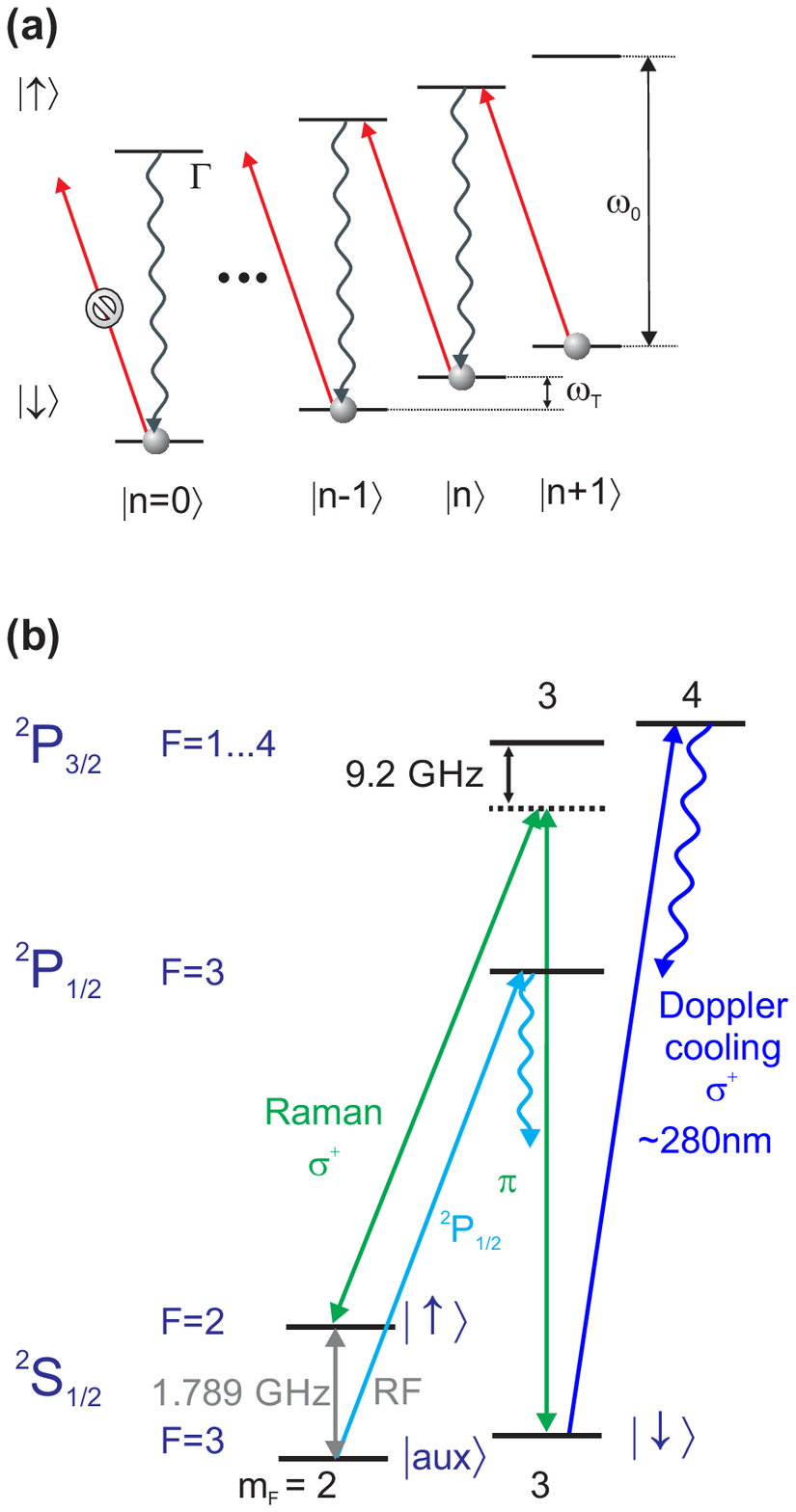}  
	\captionmacro[Principle of resolved sideband cooling and
	implementation in {\boldmath\Mgtf}.][(Color online)]{\subplotlabel{a} Resolved
	\gls{SBC} requires the linewidth $\Gamma$ of the transition to be smaller than the trap frequency $\omegaT$, so that individual transitions can be selectively driven. With the cooling laser tuned to the \nth{1} order \gls{RSB}, one quantum of motion is removed from the system in each absorption event. In the Lamb-Dicke regime, a change in motional state is suppressed upon decay back to the \downket state.
	\subplotlabel{b} Relevant level structure for SBC of a \Mgtf ion. Doppler cooling and Raman transitions are performed by coupling the
$\dsoh$ and $\dpth$ states. Compared to a previous implementation of SBC
\cite{hemmerling_single_2011}, a second laser is added for repumping the
\auxket state via the \dsoh $\leftrightarrow$ \dpoh transition. RF radiation couples the states \upket and \auxket.}  
	\label{fig:principle_sbc}    
\end{figure}      
The current work is based on previous results \cite{hemmerling_single_2011}, where
we demonstrated a  pulsed Raman sideband cooling scheme using a single laser
system for  \Mgtf. In brief, a \Mgtf ion is trapped via isotope-selective
photoionization  in a linear Paul trap with axial and radial motional
frequencies of  \SI{2.21}{MHz} and $\sim$ \SI{5}{MHz}, respectively. The
relevant levels  for \gls{SBC} of a single \Mgtf ion are shown in
\cref{fig:principle_sbc}(b). A  frequency-quadrupled fiber laser (\dpth laser in
the following)  provides the radiation for \acrlong{DC}, coherent control, and
state detection on the  \Mgtf ion. Doppler cooling on the
\bsfull$\equiv\downket$ $\leftrightarrow$  $\Ket{\dpth, F=4, m_F = 4}$
transition for \SI{400}{\mus}  yields a $\nbar\sim 10$
as the starting point for the ground state cooling sequence.  Coherent electronic and motional state manipulation is
implemented via Raman  transitions, coupling the $\downket$ and $\upket\equiv$\dsfull states.  No change in the motional state correspond to
carrier (CAR)  transitions, while addition of phonons correspond to blue
sideband (BSB)  transitions upon changing the electronic state from \downket to
\upket.  The Raman beams are generated via acousto-optic modulators from the
\dpth laser and are detuned  with respect to the \dpth state by around 9.2~GHz.
The beams have $\pi$ and  $\sigma^+$ polarization to maximize the coupling
strength and are at right angle  to each other with the difference wavevector
aligned along the axial direction  of the trap. This results in a Lamb-Dicke
parameter of $\eta=0.3$  \citep{hemmerling_single_2011} for the axial mode of
interest here. The limited  detuning of the Raman laser beams leads to
off-resonant excitations, which result  in dephasing and population loss during
coherent manipulation as further  discussed in Sec.~\ref{sec:extract_nbar}. This
effect limits the detection  fidelity of the motional ground state population
and the final $\nbar$ detectable  after cooling. 
% \cPiet{Is this really true?
% Off-resonant excitation and other  detection errors would lead to a constant
% offset, which is subtracted for  the analysis. Move to below?} 
It can be reduced
by employing a separate Raman  laser system with larger detuning.

\begin{figure}    
	\centering  
	\includegraphics[width=0.48\textwidth]{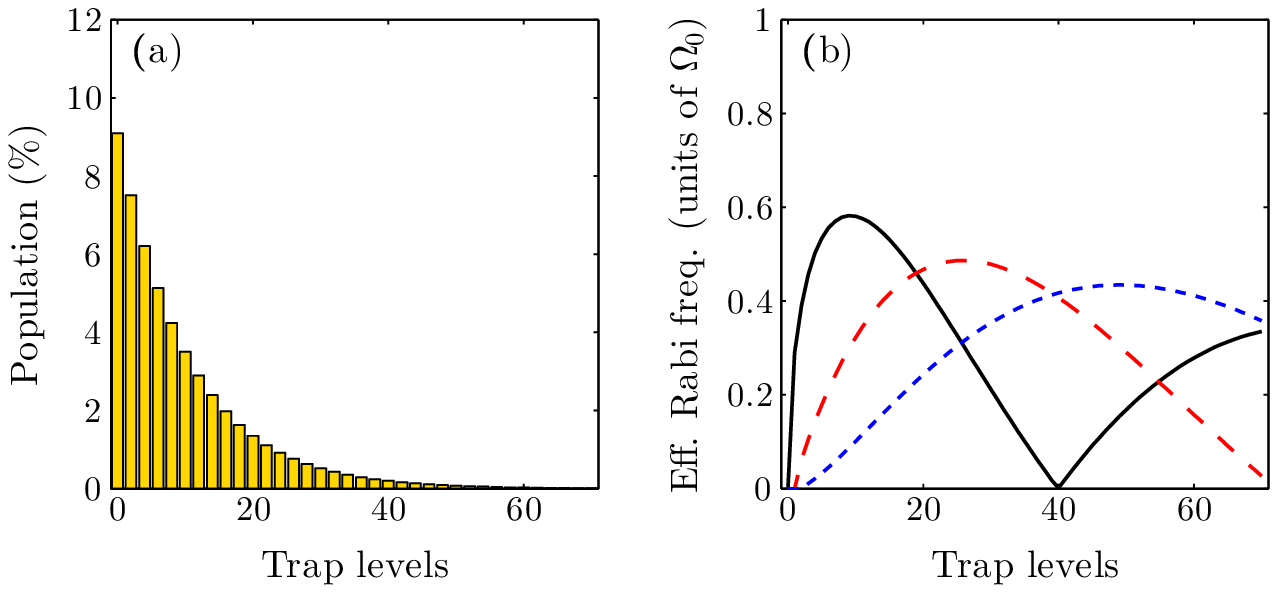}
	\captionmacro[Motional state population and effective Rabi frequencies.]{
% 	\subplotlabel{a} Distribution over
% 	the motional levels at the trap frequency of $\omegaT=\SI{2.21}{\MHz}$ and a temperature of $T =
% 	\SI{1}{mK}$. Less than 0.5\%  
% 	of the population are in the trap levels $\Ket{n>50}$. \subplotlabel{b}
	(a) Typical population distribution over motional states after Doppler cooling,
	corresponding to $\bar{n}\approx 10$. 
	%\cPiet{Not consistent! We write
	%$\bar{n}=16(5)$ in the text, while the distribution in the Figure corresponds
	% to $\bar{n}\sim 9$. How should we fix this?} 
	(b) Effective Rabi frequencies as a function of the motional state calculated with a Lamb-Dicke parameter $\eta = 0.30$ according to \cref{eqn:rabi_freq}. The effective Rabi frequencies for different RSB orders have zero points at different motional excitations. \textit{solid/dashed/dotted}: \nth{1}/\nth{2}/\nth{3} order RSB. 
% 	\cPiet{use colors: black, red, blue; change figure-part labels here and in all other figures to (a), (b),... as required by PRA style}
	\source{\url{N:/Doktorarbeit_Sync_N10451/writing/PhD_Thesis/matlab/cooling_mg25_2_22.m}}}
	\label{fig:Rabi_freq_mg25}  
\end{figure}

In the previously implemented SBC scheme \cite{hemmerling_single_2011} the Rabi
frequency $\Omega_{n',n}$ of  the Raman transition between motional state
$\ket{n}$ and $\ket{n'}$ was  calculated according to \citep{wineland_laser_1979, wineland_experimental_1998}
\begin{eqnalign}
	\label{eqn:rabi_freq}
    \Omega_{n^\prime, n} & =     \Omega_\mathrm{0}\exp\left(\frac{-\eta^2}{2}\right)\sqrt{\frac{n_{<}!}{n_{>}!}}\eta^{|n^\prime-n
    |}L_{n_{>}}^{|n^\prime-n|}(\eta^2),
\end{eqnalign}
where $\Omega_0$ is the CAR Rabi frequency, $n_{<}$ $(n_{>})$ denotes the
smaller (larger) of $n, n'$, and $L_n^a(x)$ are the generalized Laguerre polynomials. This allowed the determination of the $\pi$-time for complete population transfer on the \nth{1} and \nth{2} order RSBs (see \cref{fig:Rabi_freq_mg25}). This way, RSB $\pi$-pulses were applied to sweep the population starting from $n=40$ to the motional ground state.
%This scheme however requires the precise knowledge of the Rabi frequency of the Raman lasers to be able to predict the correct pulse length to implement a $\pi$-pulse on the \gls{RSB} transition (\cref{eqn:pulse_length_matching}). 
After each \gls{RSB} pulse, multiple repump cycles involving optical excitation of the $\upket \rightarrow \dpth$ transition and RF state inversion pulses between \auxket and \upket are necessary to transfer the population back to the \downket state, since the excited state in \dpth can also decay into the $\auxket\equiv \auxfull$ state.
%The ion can decay to the \auxket state during the repump process, so that interleaved RF pulses coupling \auxket and \upket are necessary to bring the population in \auxket back to the \downket state. 
The cooling speed of this implementation was mainly limited by the \gls{RF} coupling strength and the required number of sweeps (2-3) to reach the ground state. 

%####################################################################################
\subsection{Cooling scheme}     
%####################################################################################
\label{sec:sgl_Mg_scheme}  
In the approach described in the following, we significantly reduce the cooling time by adding a dedicated repump laser for faster repumping of the \auxket state via the $\Ket{\dpoh, F=3, m_F=3}$ state and by using RSB pulses with equal length. The experimental sequence 
%as shown in \cref{fig:seq_full_mg25} 
starts with \acrlong{DC}. For \gls{SBC}, we first apply a series of \nth{2}
order \gls{RSB} pulses followed by a series of \nth{1} order \gls{RSB} pulses of
fixed lengths. After each \gls{SBC} pulse a short ($t_\mathrm{r} =
\SI{3}{\mus}$) optical repumping pulse using the $\sigma^+$ beam of the Raman
laser tuned to resonance and the \dpoh laser is applied to clear out the
population left in $\auxket$ and $\upket$. In the experiment the actual elapsed
time between two \gls{SBC} pulses equals $t^\prime_\mathrm{r} = 5~\mu$s caused
by delays from the control electronics. In addition, we can use the \dpoh laser
and the RF coupling between \auxket and \upket to implement a quench coupling
during the Raman RSB pulses. This quench coupling opens up an additional decay
channel for the \upket state back to the ground state \downket with a
controllable decay rate, implementing a fusion between pulsed and continuous
sideband cooling. After \gls{SBC}, the population in the motionally excited
states is probed by driving a RF $\pi$-pulse to transfer all population from
\downket to \upket, followed by a \gls{STIRAP} pulse on the \nth{1} \gls{BSB}
\citep{gebert__2015}. %These combined detection pulses are used instead of a
% simple \gls{STIRAP} pulse on the \gls{RSB} because of technical limitations.                    
% With this detection scheme any population loss from the two coupled states is
% immediately indicated in the signal. In contrast when detecting with a single pulse on the \nth{1} red sideband a loss in contrast would mimic cooling and therefor lead to wrong results.
This maps motionally excited states ($n>0$) onto the state \downket, while the
motional ground state population ($n=0$) remains in the state \upket 
\footnote{Probing the population using a BSB instead of a RSB STIRAP pulse
allows for detection of population loss into other hyperfine states by reducing
the signal.}. %and completes the experimental sequence.
% %(\cref{fig:seq_full_mg25}).                  
%\begin{figure}[t!] 
%	\centering
%	\includegraphics[scale=0.8]{seq/seq_full_mg25_edited.eps}
%	\captionmacro[Full experimental sequence.]{Full experimental sequence
%	consisting of Doppler cooling, sideband band cooling, RF $\pi$-pulse, STIRAP pulse and detection pulse. \cPiet{adjust colors to make the text legible}}
%	\label{fig:seq_full_mg25}
%\end{figure}

%As an alternative \gls{SBC} scheme, we tuned the \dpoh laser to be resonant with the \upket$\leftrightarrow \Ket{\dpoh, F=3, m_F=+3}$ transition. However, this scheme fails, especially in the case of long \gls{SBC} pulses, which we attribute to off-resonant excitation from the Raman lasers leading to decay into other Zeeman sub-levels of the $F=3$ ground state manifold which are not addressed by the rf coupling and therefore lost from the cooling cycle.
% In the following we will only discuss the \gls{SBC} with the scheme described in the last paragraph.

\begin{figure}
	\centering
	\includegraphics[width=0.48\textwidth]{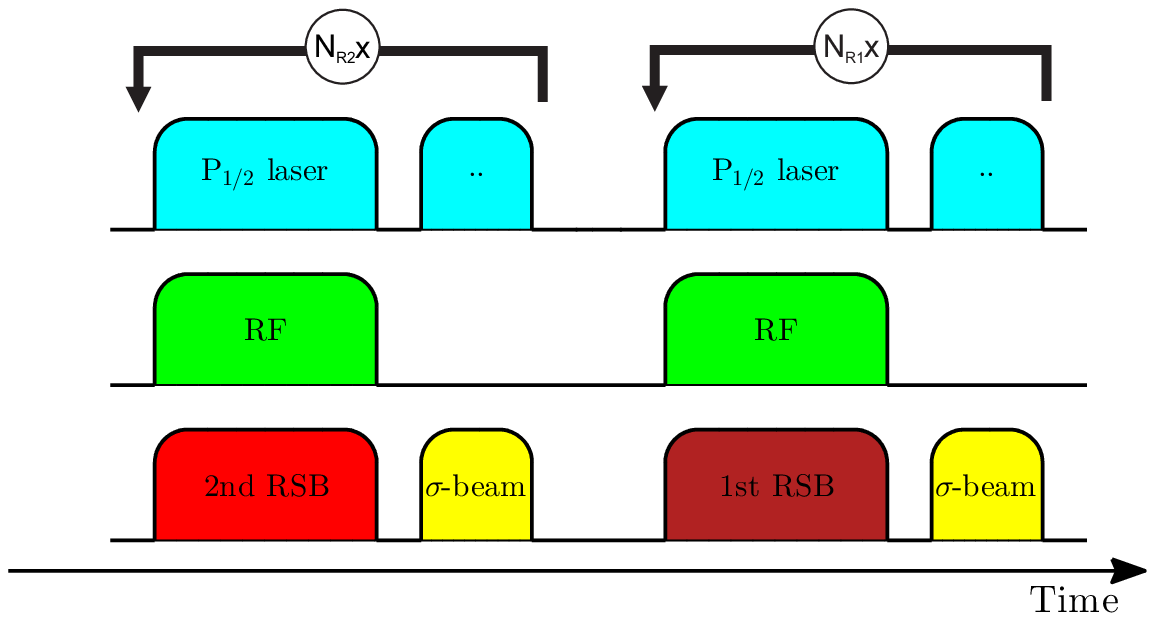}
	\captionmacro[Sequence for single ion sideband cooling.]{Sequence for \gls{SBC}
	a single \Mgtf. The sequence starts by repeating $N_\mathrm{R2}$ \nth{2} order
	RSBs, followed by $N_\mathrm{R1}$ \nth{1} order RSBs with fixed pulse lengths.
	After each \gls{SBC} pulses, repumping pulses consisting of \dpoh and $\sigma$
	beams are applied to bring the ion back to \downket state.
% 	\cPiet{extend RF line to the end; change dark blue to yellow; $\sigma$-beam is not explained}
	}
	\label{fig:seq_sbc_mg25_Scheme1}
\end{figure}

We optimize the cooling rate for a fixed total cooling time $T_c$ by changing the amount of time spent on cooling via \nth{2} and \nth{1} order RSBs, characterized by a time scaling factor $\alpha$. Depending on the pulse length of the \nth{1} (\nth{2}) order \gls{RSB} pulses $t_\mathrm{R1}$ ($t_\mathrm{R2}$), we apply $N_\mathrm{R2}$ \nth{2} order and
$N_\mathrm{R1}$ \nth{1} order \gls{RSB} pulses
\begin{equation}
\label{eqn:n_repump} 
\begin{split}
	N_\mathrm{R2} =&\ \left\lfloor\frac{\alpha T_\mathrm{c}}{t_\mathrm{R2}}\right\rfloor \\
	N_\mathrm{R1} =&\ \left\lfloor\frac{(1-\alpha) T_\mathrm{c}}{t_\mathrm{R1}}\right\rfloor,
\end{split}
\end{equation}
where the sign $\lfloor x \rfloor = \max\{ n \in \mathbb{Z},n\leq x \}$ denotes the floor function. Short padding pulses are added to keep $T_c$ fixed. The total time spent on ground state cooling during the cooling sequence as shown in \cref{fig:seq_sbc_mg25_Scheme1} is expressed as
\begin{equation}
	T_\mathrm{total} = T_\mathrm{c} + N_\mathrm{r}\cdot t^\prime_\mathrm{r},
\end{equation}
where 
$N_\mathrm{r}$ is the total number of repump pulses.  

Since we are interested in the cooling time, we investigate the dynamics of \gls{SBC} by probing the population in the motional ground state with the STIRAP sideband pulse for different \gls{SBC} times $T_\mathrm{c}$ instead of optimizing for lowest $\nbar$.
%\question{The following section for the model function and so on needs to be
% changed.} 
Assuming a constant cooling rate $W$ during the \gls{SBC} cycle, the mean occupation of the motional states decays exponentially \citep{stenholm_dynamics_1985} as 
\begin{equation} \label{eqn:nbar_decay}
	\nbar(t) = \bar{n}_\mathrm{f}(1-e^{-Wt}) + \bar{n}_\mathrm{i}e^{-Wt}
\end{equation}
where $\bar{n}_\mathrm{i}$ and $\bar{n}_\mathrm{f}$ are the initial and final mean occupation, respectively, and $W$ is the cooling rate. Assuming a thermal distribution with mean occupation of $\bar{n}(t)$ after \gls{SBC} for a duration $t$, the population in the motional ground state can be expressed as
\begin{eqnalign}
	P_\mathrm{0}(t) &= \frac{1}{1+\bar{n}_\mathrm{f}(1-e^{-Wt}) +
	\bar{n}_\mathrm{i}e^{-Wt}} \\ 
	&= \frac{1}{1+\nbar_\mathrm{f} + (\nbar_\mathrm{i}-\nbar_\mathrm{f}
	)e^{-t/T_\mathrm{0}}}\,.
\end{eqnalign}
The desired cooling time constants $T_\mathrm{0} = 1/W$ for sets of parameters ($t_\mathrm{R1}$, $t_\mathrm{R2}$, $\alpha$) are extracted by fitting data to this model function in which only one parameter is changed and $\bar{n}_\mathrm{i}$ and $\bar{n}_\mathrm{f}$ are common fit parameters. 
By choosing $T_\mathrm{c}\sim 7\cdot T_\mathrm{0}$, we ensure reaching the steady state.
%We investigate the dependence of the cooling time constant $T_\mathrm{0}$ as a function of different experimental parameters in the following section. \sourceft{D20131023}

% \begin{figure}[tbh]  
% 	\centering
% 	\includegraphics[width=12.5cm]{mg25/csbc_fit_check_tsf_mit_repump_tex.eps}
% 	\caption{Example for fitting the model to the experimental data.
% 	\source{D20131023}}  
% 	\label{fig:csbc_fit_check_tsf_mit_repump}  
% \end{figure}        

% \section*{Appendix}
\subsection{Optimization of experimental parameters}  
\label{sec:results_mg25}

Using the procedure described in the previous section, we vary the pulse length of the \gls{RSB} pulses, the time scaling factor $\alpha$ and the optical power of the \dpoh repump laser to optimize the cooling rate. Furthermore, we explore the transition between pulsed and quench cooling by adding the \dpoh repump laser and \gls{RF} coupling between the \auxket and \upket states during the \gls{RSB} pulses. A numerical simulation based on optical Bloch equations described in \cref{app:numerical_simulation_SBC} supports our findings quantitatively.

 \begin{figure}
	\centering
	\includegraphics[width=0.48\textwidth]{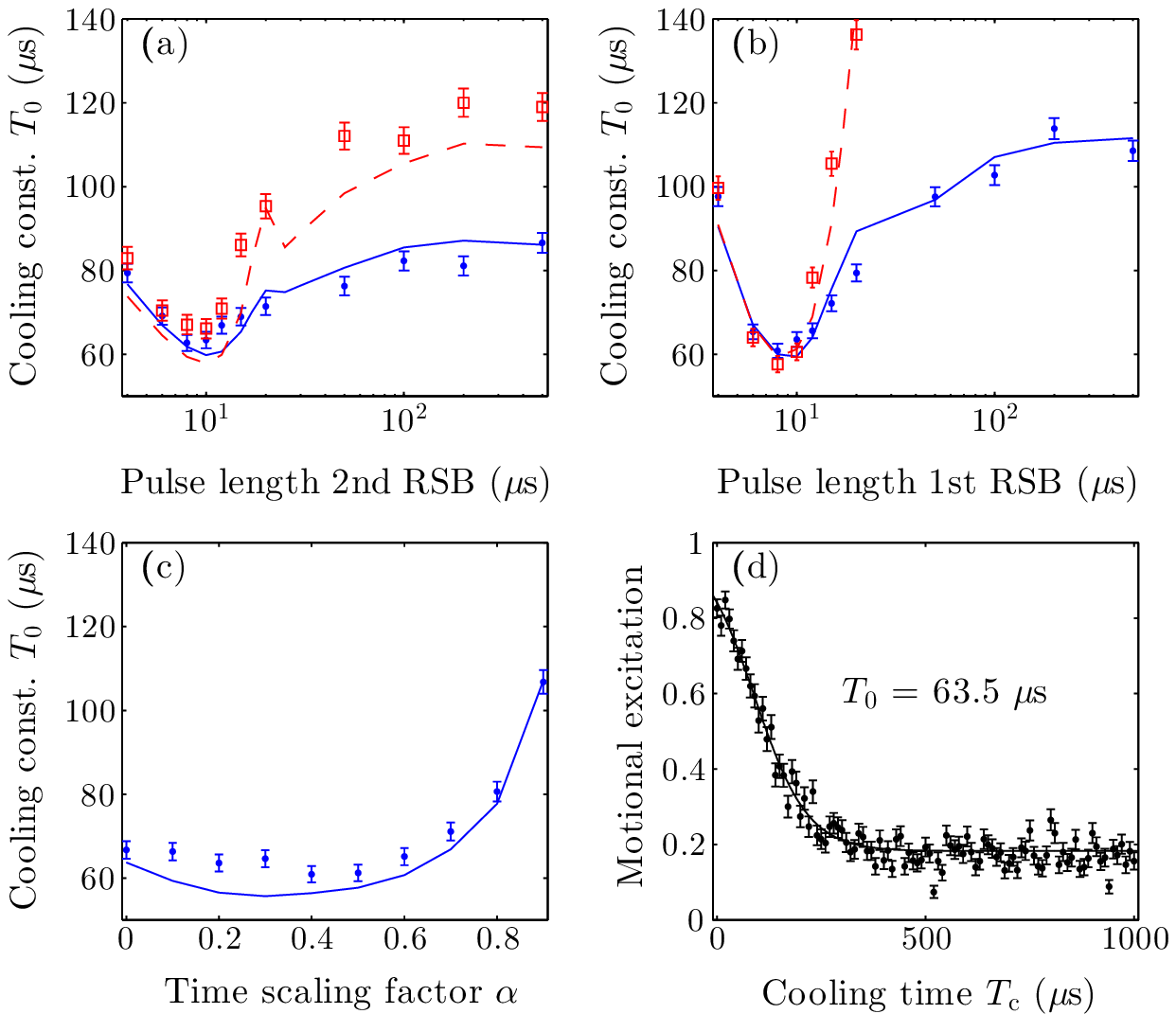}
	\captionmacro[Dependence of the cooling time constants on
	experimental parameters.]{
	Cooling time constants as a function of the pulse lengths of the \subplotlabel{a} \nth{2} order and \subplotlabel{b} \nth{1} order \gls{RSB} pulses with $\alpha = 0.5$ while fixing the non-scanned pulse length near its optimal value.
	%$t_\mathrm{R2} = \SI{10}{\mus}$, $t_\mathrm{R1} = \SI{8}{\mus}$
	\textit{blue point (red square)}:
	Experimentally determined cooling constants with the quench coupling on (off) during the RSB pulse. The quench coupling induces an effective decay rate of \SI{42}{ms^{-1}} from the \upket state to the \downket state.
	\subplotlabel{c}: 
	Cooling time constant as a function of the time scaling factor with $t_\mathrm{R2} = \SI{10}{\mus}$ and $t_\mathrm{R1} = \SI{10}{\mus}$.
	The lines in (a)-(c) are the result of Master equation simulations using the
	experimental parameters. \subplotlabel{d} Residual motional excitation as a
	function of the total cooling time $T_\mathrm{c}$. The line is the fit to
	the experimental data, which gives the cooling constant $T_\mathrm{0}$. 
% 	\cPiet{Replace figure part (d) with single T0 measurement?.}
	\source{D20131023 for a-c.} 
	%\source{D20131126 for d.} 
	\source{\url{Mg25_CSBC/matlab/sinc_func.m} for d.} 
	\source{Figure generated with
	\url{difcos/daily_evaluation/20131023/sbc_time.m}}}
	\label{fig:cooling_constant_mit_repump}    
\end{figure}    

% \begin{figure}[tbh]  
% 	\centering
% 	\includegraphics{theory/sinc_tex.eps}
% 	%\input{mg25/cooling_constant_2rsb_mit_repump.tikz}
% 	\caption{Optimal pulse length for \gls{SBC} pulses. \textit{Solid black line}: Rabi
% 	oscillation with its maximum reached at $T_\mathrm{\pi}$. \textit{Dashed blue
% 	line}: Mean excitation rate from Rabi excitation achieves its maximum at about
% 	$0.73\,T_\mathrm{\pi}$.
% 	\sourcedet{Mg25_CSBC/matlab/sinc_func.m}}  
% 	\label{fig:theory_sinc}
% \end{figure}

%###
\subsubsection{Pulse length optimization}  
%###
We first optimize the cooling time constant as a function of the pulse length
for the \nth{1} and \nth{2} order  \gls{RSB} pulses. For each parameter set, a
scan  of the \gls{SBC} time $T_\mathrm{c}$ is performed, which allows us to
derive the  corresponding cooling time constant as shown in
\cref{fig:cooling_constant_mit_repump}(d). 
%  \cPiet{Do we want to implement this?}.
  In \cref{fig:cooling_constant_mit_repump}(a) and \ref{fig:cooling_constant_mit_repump}(b), the experimentally determined cooling time constants $T_\mathrm{0}$ are plotted against the pulse lengths. The shortest $T_\mathrm{0}$ is obtained for \nth{1} and \nth{2} order \gls{RSB} pulse lengths of around \SI{10}{\mus}.
%The Rabi frequency for transitions changing an initial state $n$ 
The experimentally determined $\pi$-time for a \gls{RSB} pulse starting at $\ket{n=1}$ is $T_\mathrm{\pi} \approx \SI{16}{\mus}$. Averaging over the $\pi$-times for \nth{1} and \nth{2} order RSBs according to \cref{eqn:rabi_freq}, weighted by the thermal occupation of the initial states, limited for each sideband to the states where the coupling strength of the respective sideband dominates, gives for both sidebands average $\pi$-times of $\sim$\SI{10}{\mus} in agreement with the experimentally optimized values. The fact that a constant $\pi$-time is sufficient for efficient cooling is a consequence of the relatively small variation of the Rabi frequencies in the relevant range of motional levels as can be seen from \cref{fig:Rabi_freq_mg25}.

In standard pulsed Raman SBC, the cooling time constant strongly rises upon deviation from the optimum pulse lengths [red points and dashed lines in \cref{fig:cooling_constant_mit_repump}(a) and (b)].
%More precisely, the mean excitation rate decreases for short \gls{SBC} pulses as illustrated in \cref{fig:cooling_constant_mit_repump}. 
This dependence becomes much weaker for long pulses [blue points and solid lines in \cref{fig:cooling_constant_mit_repump}(a) and (b)] by applying the quench coupling during the RSB pulses which opens a decay channel for the $\upket$ state. In the experiment, this decay rate was adjusted to about \SI{42}{ms^{-1}} or a decay constant of about \SI{24}{\mus} to achieve the highest cooling rate. At this decay rate, the effect of repumping is negligible for pulses shorter than the optimum pulse length and the scheme is equivalent to standard pulsed \gls{SBC}. For \gls{RSB} pulses longer than the optimum pulse length, the ion in the $\upket$ state decays back to the $\downket$ state through the new channel. The ion cycles between the $\downket$ and the $\upket$ states and therefore more than one phonon can be removed within a single RSB pulse. The scheme thus becomes equivalent to continuous \gls{SBC}. This quasi-continuous cooling scheme is insensitive to the exact pulse length of the \gls{RSB} pulses and provides high robustness against intensity/pointing fluctuation of the Raman lasers.

%\question{Why is the total duration for long pulse length longer?}

% \begin{figure}[tbh]
% 	\centering
% 	\includegraphics{mg25/comparison_with_without_repumper_tex.eps}
% 	\caption{Effect of repumper during Raman sideband cooling
% 	in \schemeA. \source{D20131023}}
% 	\label{fig:comparison_with_without_repumper}
% \end{figure}

% \begin{figure}[tbh]      
% 	\centering
% 	\includegraphics[width=12cm]{mg25/cooling_constant_2rsb_mit_repump_tex.eps}  
% 	%\input{mg25/cooling_constant_2rsb_mit_repump.tikz}   
% 	\caption{Dependence of cooling constant on pusle length of \nth{2} RSB in
% 	\schemeA. \source{D20131023}}
% 	\label{fig:cooling_constant_2rsb_mit_repump}
% \end{figure}
% \begin{figure}[tbh]
% 	\centering
% 	\includegraphics[width=12cm]{mg25/cooling_constant_1rsb_mit_repump_tex.eps}
% 	\caption{Dependence of cooling constant on pusle length of \nth{1} RSB in
% 	\schemeA. \source{D20131023}}
% 	\label{fig:cooling_constant_1rsb_mit_repump}
% \end{figure}
    
%###
\subsubsection{Time scaling factor optimization}
%###
The time scaling factor $\alpha$ as defined in \cref{eqn:n_repump}  distributes
the \gls{SBC} time $T_\mathrm{c}$ into the time spent on the \nth{2}  order
\gls{RSB} $T_\mathrm{R2} = \alpha T_\mathrm{c}$ and on the \nth{1} order 
\gls{RSB} $T_\mathrm{R1} = (1-\alpha)T_\mathrm{c}$. Due to the dependence  of
the Rabi frequency on the motional quantum number (see \cref{fig:Rabi_freq_mg25}
and \cref{eqn:rabi_freq}), the \nth{2} order \gls{RSB} pulses are more efficient
in cooling the  population in motional states $n>20$ at the starting stage of the
\gls{SBC} cycle.  For the lower motional states, the \nth{1} order \gls{RSB}
pulses become more efficient,  so that an $\alpha$ that is too large also increases the
\gls{SBC} cooling  time constant (\cref{fig:cooling_constant_mit_repump}{c}). 
% \cPiet{I think we should be more specific what we mean with ``more efficient''. 
% If we take the $\pi$-time per phonon as the figure of merit, we would have to 
% weight the Rabi frequencies according to the sideband order, shifting the
% optimum  transition between 1st and 2nd order to lower $n$. This may be
% particularly  important for the higher order cooling discussed belowe.} 
Since the variation of \nth{1} and \nth{2} order RSB Rabi frequencies with $n$ are very slow, the cooling time spent on either order only significantly influences the cooling time constant for the extreme cases of $\alpha\rightarrow 0$ and $\alpha\rightarrow 1$ and remains otherwise flat.

\subsection{Cooling results}
\label{sec:extract_nbar}
\begin{figure}
	\centering
	\includegraphics[width=0.48\textwidth]{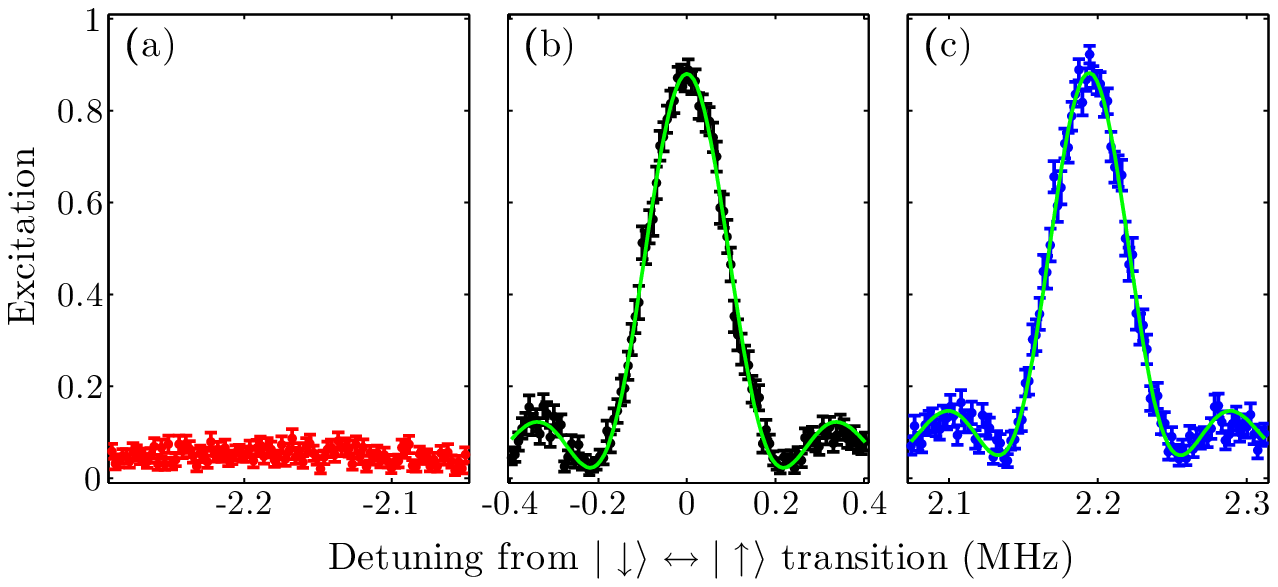}
	\captionmacro[Rabi excitation of {\boldmath\Mgtf} after sideband  
	cooling.]{\subplotlabel{a-c} Frequency scans of Raman transitions over \nth{1} \gls{RSB} (\textit{red}), carrier (\textit{black}) and \nth{1} \gls{BSB} (\textit{blue}) transitions after \gls{SBC} of a single \Mgtf.
% 	\cPiet{remove figure parts d and e}
	%\subplotlabel{d} Raman Rabi flop on carrier \subplotlabel{e} Raman Rabi flops on blue and red sideband after \gls{SBC} performed on a single \Mgtf. The \textit{solid green lines} are fits to the data. The \textit{dashed orange line} indicates the boundary set by  off-resonant excitation from the dark state and the \textit{dashed black line} indicates the boundary set by   off-resonant excitation from the bright state.
	\source{D20131104}}
	\label{fig:freq_mg25_Scheme1}     
\end{figure}
         
% \begin{figure}[b!] 
% 	\centering
% 	\includegraphics{mg25/flop_mg25_pi_detection_tex.eps}
% 	\captionmacro[Rabi oscillation from the motional ground
% 	state.]{\subplotlabel{a} Raman Rabi flop on Carrier \subplotlabel{b} Raman Rabi
% 	flops on blue and red SB after \gls{SBC} performed on a single \Mgtf. The dashed orange line indicates the boundary set by the off-resonant excitation from the dark state and the dashed black line
% 	indicates the boundary set by the off-resonant excitation from the bright
% 	state.
% 	\source{D20131104}}
% 	\label{fig:flop_mg25_Scheme1}  
% \end{figure}
The final \nbar after \gls{SBC} is extracted from the red and blue sideband excitation. Assuming that the ion stays in a thermal distribution after \gls{SBC}, the ratio of excitations on \nth{1} red and blue sideband fulfills the following relation \citep{turchette_heating_2000}   
\begin{equation}
	Q\coloneqq\frac{\rho^\mathrm{rsb}(t)}{\rho^\mathrm{bsb}(t)}=\frac{\nbar}{1+\nbar},
\end{equation}    
where $\rho^\mathrm{rsb}(t)$ and $\rho^\mathrm{bsb}(t)$ are the excitation probabilities on the \nth{1} RSB and BSB at time $t$, respectively. For this analysis we subtracted a constant signal background from the data. With the \gls{SBC} sequence described in \cref{sec:sgl_Mg_scheme}, we achieve $\nbar = \rho^\mathrm{rsb}/(\rho^\mathrm{bsb}-\rho^\mathrm{rsb}) \approx 0.01(2)$ as shown in \cref{fig:freq_mg25_Scheme1}(a)-(c) after a cooling time of $T_\mathrm{c} = \SI{500}{\mus}$. This represents a reduction in cooling time by more than a factor of 30 compared to reference \cite{hemmerling_single_2011}. %The cooling sequence consists in total of only 50 pulses \cPiet{this is my estimate, is the number correct?} equally distributed between \nth{2} and \nth{1} order RSB transitions to cool the motional state population shown in \cref{fig:Rabi_freq_mg25}.

\section{Sympathetic ground-state cooling of a molecular ion}   
\label{sec:sbc_two_ion_crystal}     

In this Section we adapt the quasi-continuous cooling scheme for sympathetic cooling of a molecular ion using \Mgtf as the cooling ion species. 
\subsection{Loading and sympathetic Doppler cooling of {\boldmath\Mghd}}  
A two-ion crystal consisting of a \Mgtf and a \Mghd ion is prepared by
isotope-selective photo-ionization loading \cite{kjaergaard_isotope_2000} of a
pair of \Mgtf and \Mgtv ions with the Doppler cooling laser tuned near the
resonance of the \Mgtf cooling transition. Then, the laser is tuned near to the
cooling resonance of \Mgtv and hydrogen gas is leaked into the vacuum system
through a leak valve, increasing the pressure up to $\sim 5\times 10^{-9}$~mbar.
After a photo-chemical reaction during which the excited \Mgtv ion reacts with a
hydrogen molecule to form \Mghd \cite{molhave_formation_2000}, the fluorescence
of the \Mgtv ion vanishes. After closing the leak valve and tuning the laser
back to the \Mgtf resonance, cooling commences and the mass of the dark ion is
determined via mass spectrometry using parametric heating of the two-ion crystal
\cite{drewsen_nondestructive_2004}. Since the mass of the two ions is almost
identical, we do not expect significant deviations from Doppler cooling
temperature, even in the presence of additional heating
\citep{wubbena_sympathetic_2012}.                      

\subsection{Two-mode sympathetic ground-state cooling sequence}
\begin{figure}
	\centering
	\includegraphics[width=0.48\textwidth]{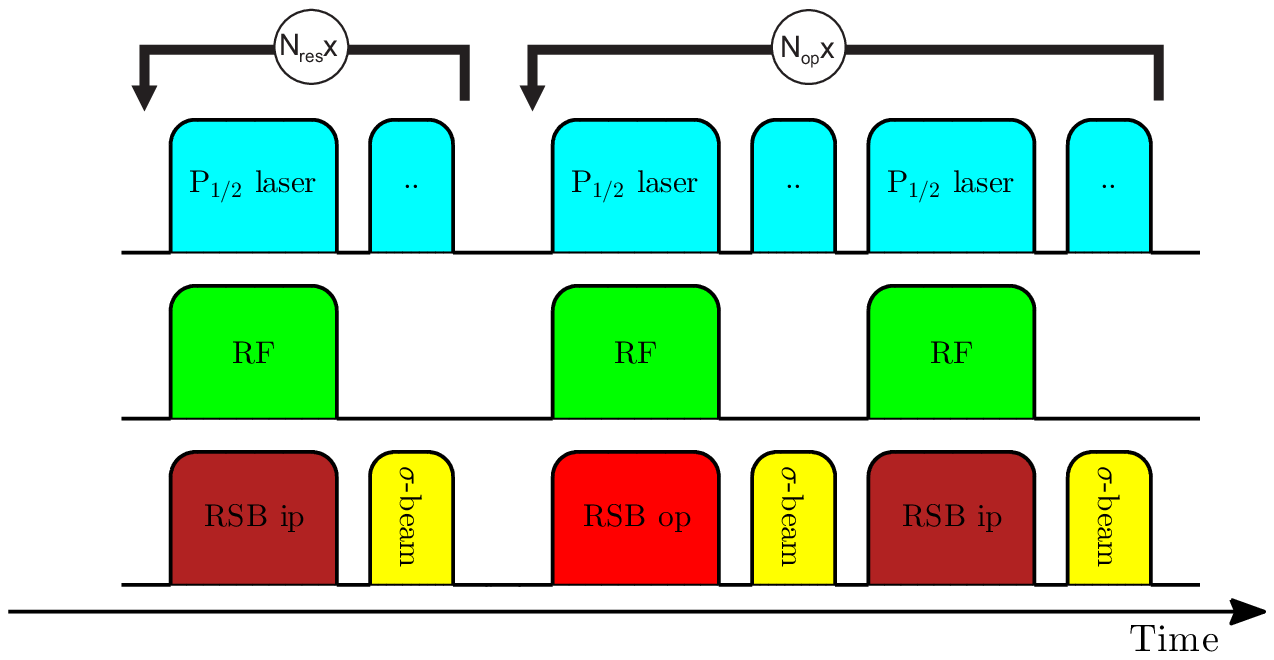}
	\captionmacro[Sequence for sympathetic sideband cooling of a two-ion crystal.]{The \acrlong{mode:ip} and \acrlong{mode:op} axial modes of a \Mgtf/\Mghd 2-ion crystal are cooled in an interleaved fashion to ensure simultaneous cooling. 
% 	\cPiet{change adjust color; extend line}
}  
\label{fig:seq_sbc_molecule}
\end{figure}

For sympathetic ground state cooling of the molecular ion we use a slightly modified pulse sequence compared to \cref{sec:sgl_Mg_scheme}. In contrast to the case of a single \Mgtf, the motion of the ions is described by two modes along the axial direction, the \gls{mode:ip} and the \gls{mode:op} mode with secular frequencies of $\omegaT^\mathrm{ip} = 2\pi\times\SI{2.21}{\MHz}$ and $\omegaT^\mathrm{op} = 2\pi\times \SI{3.85}{\MHz}$. The Lamb-Dicke parameters for the coupling of the Raman lasers to the \Mgtf ion are $\eta_\mathrm{ip} = 0.21$ and $\eta_\mathrm{op} = 0.16$. With these Lamb-Dicke parameters, the effective Rabi frequencies show no zero-crossings over the range of trap levels with significant population after \acrlong{DC}, so no higher order \gls{RSB} pulses are necessary. However, with Lamb-Dicke factors as large as these, cooling of a single mode is not sufficient to enable high fidelity operations involving only this mode. The other mode acts as a spectator mode which modifies the effective Rabi frequency of the mode of interest depending on its motional state \cite{wineland_experimental_1998}. 
%One of the consequences of this mode cross-talk is that even in principle, no perfect $\pi$ pulse on the RSB can be implemented for a specific initial motional state of a single mode. 
Therefore, we have implemented an interleaved pulse sequence for \gls{SBC} both axial modes simultaneously
%employing only \nth{1} order \gls{RSB} pulses 
as shown in \cref{fig:seq_sbc_molecule}. Depending on the time scaling factor $\alpha^\prime$, which in this case distributes the total \gls{SBC} time $T_\mathrm{c}$ into time spent for  \gls{RSB}  on the ip mode $(1-\alpha^\prime)T_\mathrm{c}$ and time for \gls{RSB} on the op mode $\alpha^\prime T_\mathrm{c}$, we apply
\begin{align}
	N_\mathrm{ip} =&\ \left\lfloor \frac{(1-\alpha^\prime)T_\mathrm{c}}{t_\mathrm{ip}}
	\right\rfloor
	\\  
	N_\mathrm{op} =&\ \left\lfloor \frac{\alpha^\prime T_\mathrm{c}}{t_\mathrm{op}}   
	\right\rfloor
\end{align}
pulses on the \gls{mode:ip} and the \gls{mode:op} mode, respectively. Here $t_\mathrm{ip}$, $t_\mathrm{op}$ are the pulse lengths of the \gls{RSB} pulses on the ip and the op mode. In the case $N_\mathrm{ip} > N_\mathrm{op}$ ($N_\mathrm{ip} < N_\mathrm{op}$) we start the \gls{SBC} cycle with $N_\mathrm{res} = |N_\mathrm{ip}-N_\mathrm{op}|$ pulses on the \gls{mode:ip} (\gls{mode:op}) mode, followed by $N_\mathrm{op}$ ($N_\mathrm{ip}$) pulses on the \gls{mode:ip} and the \gls{mode:op} mode in an interleaved fashion. %At the end of the \gls{SBC} sequence, short pulses are applied to compensate the residual time. 
After every single \gls{SBC} pulse, a repumping pulse as described above is applied to clear out the \upket and \auxket states in addition to the quench coupling present also during the RSB pulses. Optimizations similar to the case of a single \Mgtf are performed to minimize the total duration of \gls{SBC}.

%###
\subsection{Pulse length optimization}
\begin{figure}
	\centering
	\includegraphics[width=0.48\textwidth]{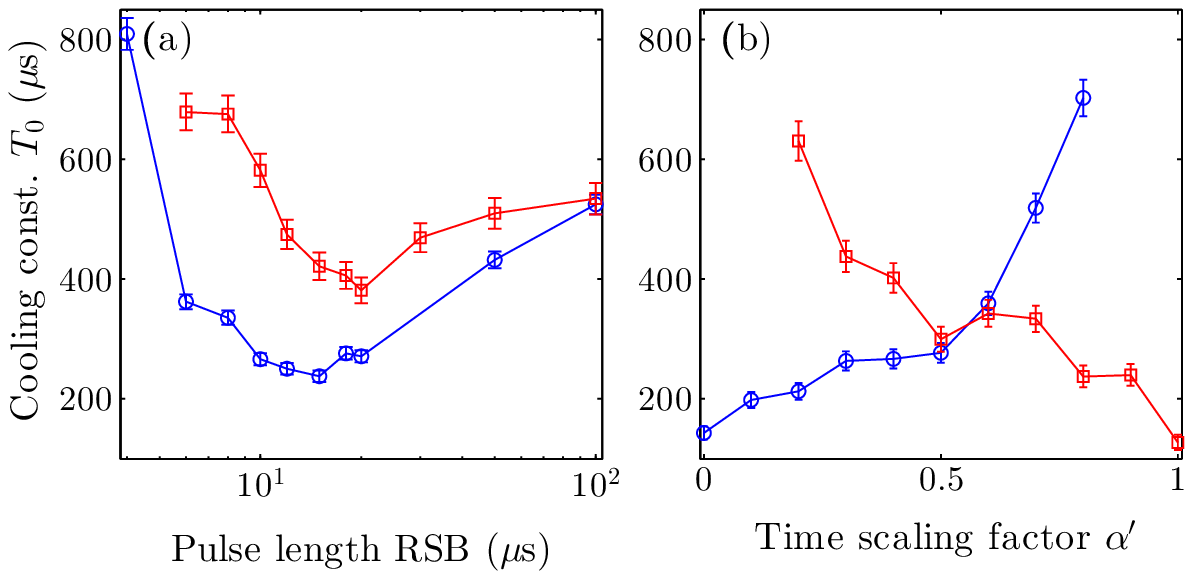}
	\captionmacro[Dependence of the cooling time constant on experimental
	parameters for sympathetic sideband cooling a two-ion
	crystal.]{\subplotlabel{a} The cooling time constant as a function of the
	optimal pulse length of the Raman \gls{RSB} pulses on the ip (\textit{blue
	circles}) and op (\textit{red squares}) mode. \subplotlabel{b} The cooling
	constant as a function of the optimal time scaling factor $\alpha^\prime$. The lines
	connect the points and are guides to the eye.                
% 	The frequency of
% 	the STIRAP detection pulse is tuned to the \nth{1} \gls{BSB} of the
% 	\gls{mode:ip} (\textit{blue circles}) and \gls{mode:op} (\textit{red squares})
% 	mode, respectively, for detecting the corresponding mode.
	\source{D20131031/D20131101, figure generated with
	\url{/difcos/daily_evaluation/20131101/sbc_time_paper.m}}}
	\label{fig:cooling_constant_molecule}    
\end{figure}

%###
The length of the \gls{RSB} pulse on the \gls{mode:ip} (\gls{mode:op}) mode is
optimized by fixing the  other mode's RSB pulse length.
%while scanning the RSB pulse length of the mode of interest.
As described in \cref{sec:sgl_Mg_scheme}, the cooling time constant is  derived
from the motional ground state population measured by performing a STIRAP pulse
resonant  with the \nth{1} \gls{BSB} of either mode for different cooling times
$T_c$. We spent  an equal amount of time for cooling on both modes ($\alpha^\prime=0.5$)
and ensure near  steady-state conditions by performing cooling up to $T_c\approx
2.5~$ms.
% \cPiet{This is not really significantly beyond steady state, but rather
% corresponds  to the optimized cooling time or have I misinterpreted anything?} 
The highest cooling rate is observed with a pulse length of 15~$\mu$s
(20~$\mu$s) for the \gls{RSB} on the \gls{mode:ip} (\gls{mode:op}) mode as shown
in \cref{fig:cooling_constant_molecule}a. With a $\pi$-time of
$T_\mathrm{\pi}^\mathrm{(ip)} \approx \SI{21}{\mus}$ and
$T_\mathrm{\pi}^\mathrm{(op)} \approx \SI{26.5}{\mus}$ for the \nth{1} \gls{RSB}
starting in the $\ket{n=1}$ state, the Rabi frequency according to
\cref{eqn:rabi_freq} averaged over the thermal occupation of motional states
results in a mean $\pi$-time of 14~$\mu$s and 25~$\mu$s for the ip and op mode,
respectively. 
% agrees with the optimized values.          
% \cPiet{I have not checked this yet!! Please do!}
 
% \cPiet{Why is the red curve higher than the blue?}

As in the single-ion case discussed in \cref{sec:results_mg25}, the increase in the cooling time constant for pulse lengths longer than the optimum pulse length is rather mild as a consequence of the quench coupling during RSB pulses.

\begin{figure*}
 	\centering
 	\includegraphics[width=0.7\textwidth]{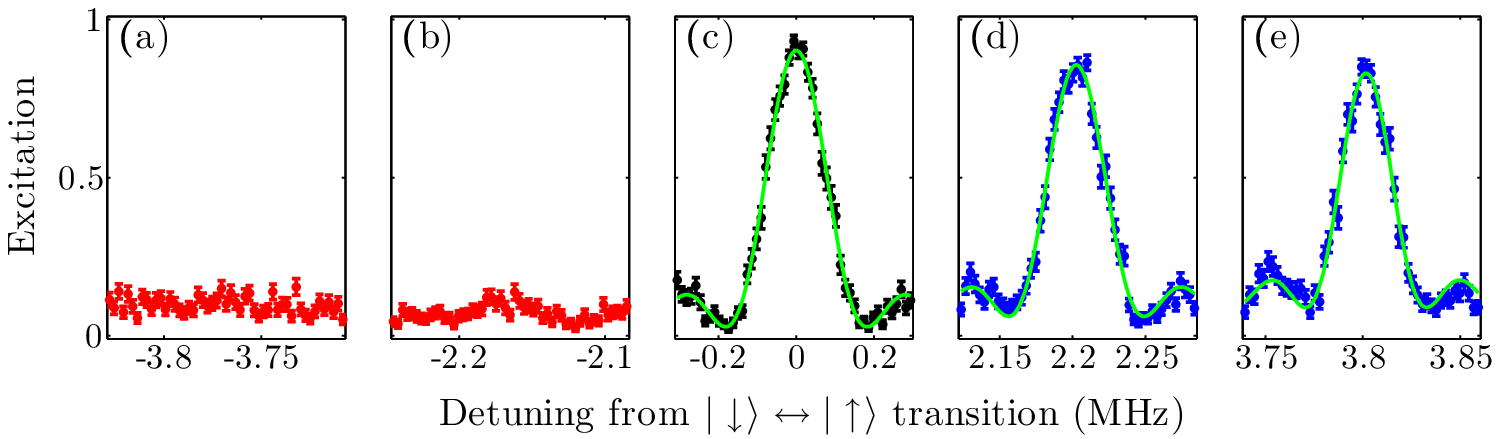}
	\captionmacro[Rabi excitation of {\boldmath\Mgtf} after 2-mode SBC a
	{\boldmath\Mghd/\Mgtf} crystal.]{Frequency scans of Raman  
	transitions over the
	\nth{1} \gls{RSB} (\textit{red}) of the \subplotlabel{a} \gls{mode:op} and \subplotlabel{b} \gls{mode:ip} modes, \subplotlabel{c} the carrier (\textit{black}), and the \nth{1} \gls{BSB} (\textit{blue}) \subplotlabel{d} \gls{mode:op} and \subplotlabel{d} \gls{mode:op} modes after \gls{SBC} both modes on \Mgtf.}
 	\label{fig:freq_molecule_Scheme1}
\end{figure*}
 
%###
\subsection{Time scaling factor optimization}
%###
The time scaling factor in the \gls{SBC} sequence for cooling a two-ion crystal divides the total cooling time $T_\mathrm{c}$ into the time spent on cooling the \gls{mode:ip} mode and the \gls{mode:op} mode for the optimized pulse lengths derived from the measurements in the previous section.
%Setting the frequency of the STIRAP pulse to  be resonant with the \nth{1} \gls{BSB} of the \gls{mode:ip} (\gls{mode:op}) mode, 
The measured \gls{SBC} cooling time constants for the two modes as a function of
$\alpha^\prime$ are shown   in \cref{fig:cooling_constant_molecule}b. Although an $\alpha^\prime$
that is too small (large) accelerates the cooling of the \gls{mode:ip}
(\gls{mode:op}) mode, the cooling time of the opposite mode increases.
% \cPiet{Does this actually make sense? I would expect the cooling rate per pulse
% in this case (in contrast to 1st and 2nd order RSB cooling) to stay constant.
% What looks like a change in rate could rather be a change in the final n. What's
% your thought on this?} 
The optimal value for the time scaling factor
$\alpha^\prime\approx 0.5$ is determined by the point where \gls{SBC} of both the
\gls{mode:ip} and \gls{mode:op} mode are achieved with the shortest total
duration of the \gls{SBC} cycle. Interestingly, this corresponds to a different
number of RSB pulses for the two modes.
%Deviating from $\alpha^\prime=0.5$ allows fine-tuning the final population in the two modes under limited cooling time according to the requirements of the experiment to be performed after ground state cooling. 
%###
\subsection{Cooling results}  
%###
\cref{fig:freq_molecule_Scheme1} shows frequency scans of the carrier and both
RSB and BSB transitions after \gls{SBC} the \gls{mode:ip} and \gls{mode:op}
modes along the axial direction\sourceft{D20131101} with optimized parameters.
The final \nbar is determined from the red and blue sideband excitations of each
mode as described in \cref{sec:extract_nbar}. As before, the offset from
off-resonant Raman excitations has been subtracted for the analysis. We reach a
$\nbar_\mathrm{ip}=0.06(3)$ for the ip mode and $\nbar_\mathrm{op}=0.03(3)$ for
the op mode after a total cooling time $T_\mathrm{total}\sim 2.5$~ms. Compared
to an extension of the SBC scheme described in reference
\cite{hemmerling_single_2011} to 2-mode ground state cooling, the optimized new
scheme reduces the cooling time by a factor of 8. 
% \cPiet{Anything we could add
% here?? Limitations of final n-bar?}            
%(\cref{fig:freq_molecule_Scheme1} and
%\ref{fig:flop_molecule_Scheme1})

% \begin{figure}[tbh]
% 	\centering    
% 	\includegraphics[width=\textwidth]{molecule/flop_molecule_Scheme1_tex.eps}
% 	\caption{Raman Rabi flops after \gls{SBC} with \schemeA performed on
% 	molecule. \source{D20131101}}
% 	\label{fig:flop_molecule_Scheme1}
% \end{figure}
   
%In conclusion, the \gls{SBC} scheme developed here requires a shorter total duration compared to the old scheme published in \cite{hemmerling_single_2011} and a high stability against fluctuations of laser intensity and thus simplifies the optimization procedure in the experiment.
          
%####################################################################################
\section{Sideband cooling beyond the Lamb-Dicke regime}
\label{sec:low_trap_frequency}
%####################################################################################
% \cPiet{From the logic of the paper, this could also be placed directly after the single-ion section before th molecules. What do you think?}
For some applications large Lamb-Dicke factors are desirable, since they enhance
the sensitivity of the ion's motion to small forces. This is the case for photon
recoil spectroscopy \citep{wan_precision_2014} or the non-destructive internal state detection of a molecular ion using oscillating dipole forces \citep{schmidt_spectroscopy_2006, vogelius_probabilistic_2006, koelemeij_blackbody_2007}.
%For a single ion, SBC to the ground state at high motional frequencies followed by adiabatically lowering of the trapping potential has been demonstrated \cite{poulsen_adiabatic_2012}. 
In the following we extend the single-ion ground state cooling scheme presented in \cref{sec:sbc_of_single_ion} to enable efficient ground state cooling outside the Lamb-Dicke regime, which is readily extended to the multi-ion case discussed \cref{sec:sbc_two_ion_crystal}.
%for a Lamb-Dicke factor as large as $\eta=0.45$. 

\begin{figure}[b]
	\centering
	\includegraphics[width=0.48\textwidth]{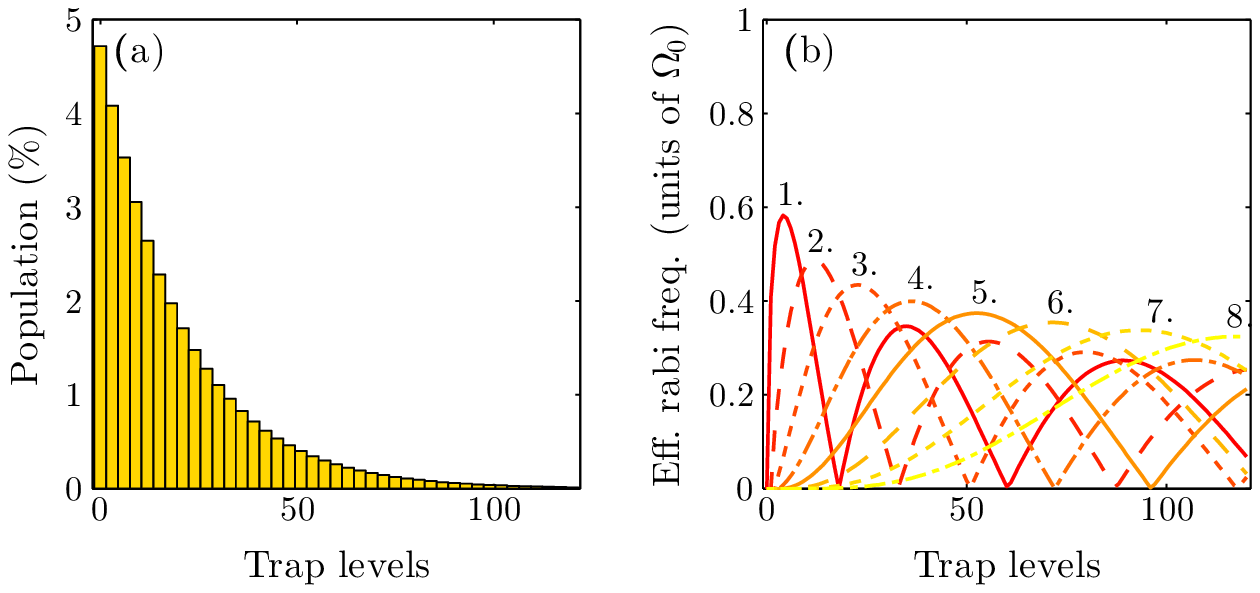}
	\captionmacro[Motional state population and effective Rabi frequency for small trap frequency.]{\subplotlabel{a}
	Distribution of the motional levels at a motional frequency of $\omegaT=\SI{1}{\MHz}$ and a temperature of $T = \SI{1}{mK}$ corresponding to the \acrlong{DC} limit of the \Mgtf ion. Less than 0.3\% of the population are in the trap levels $\Ket{n>120}$. \subplotlabel{b} 
	Effective Rabi frequencies at $\eta = 0.45$ for different \gls{RSB} orders are shown as a function of the trap levels. For ions located in  higher trap levels, \gls{RSB} pulses of higher orders $\beta(n)$ are more efficient for cooling.
	\source{See script \url{cooling_trapped_ions.m}}}  
	\label{fig:Rabi_freq_low_trap_freq}
\end{figure}
\begin{figure}
	\centering
	\includegraphics[width=0.48\textwidth]{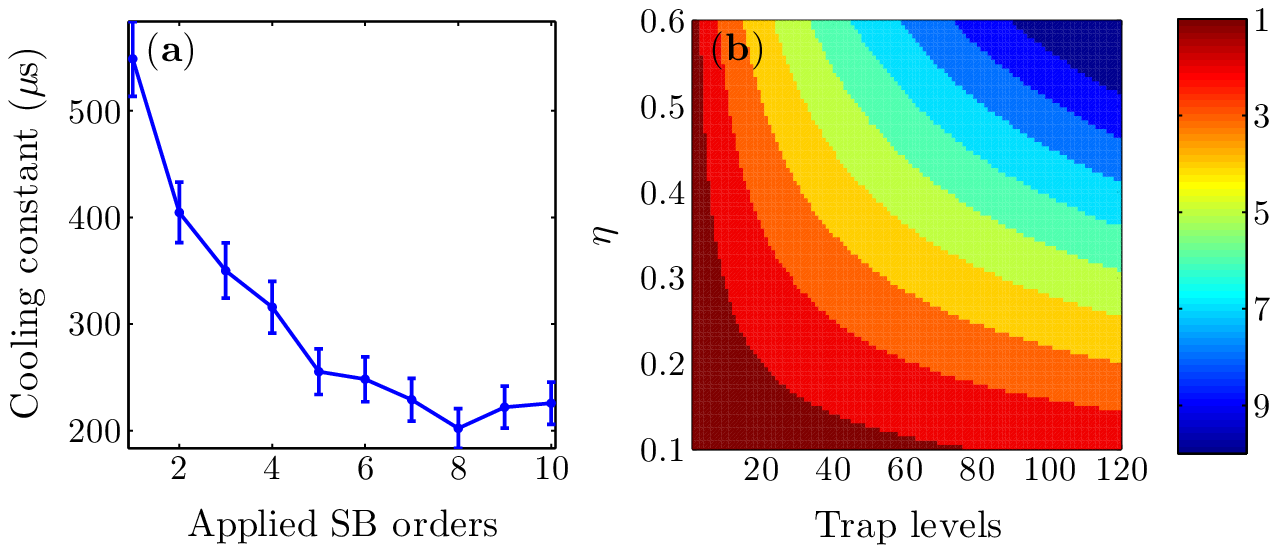}  
	\captionmacro[Determination of the maximum sideband order.]{\subplotlabel{a}
	The cooling time constant as a function of applied sideband orders. The optimal number for the sideband orders determined experimentally confirms the prediction shown in \cref{fig:Rabi_freq_low_trap_freq}(b). \subplotlabel{b} \gls{RSB} orders with highest effective cooling rate as a function of the trap level and the Lamb-Dicke parameter. With known Lamb-Dicke parameter and motional distribution a corresponding \gls{SBC} sequence can be adopted.
	\source{For a
	\url{D20131106/CSBC_time_paper.m},
	for b \url{project/cooling_trapped_ions/which_order_SB.m}}}
	\label{fig:cooling_constant_max_order}      
\end{figure}

 
At a trap frequency of \SI{1}{MHz} and a temperature of \SI{1}{mK} theoretically achievable with \acrlong{DC} a \Mgtf ion, states up to $n\sim 120$ are significantly populated, leaving less than 0.3\% population in levels $n>120$ (\cref{fig:Rabi_freq_low_trap_freq}{a}). At this trap frequency, the Lamb-Dicke parameter in our system is $\eta=0.45$. The effective Rabi frequency for \gls{RSB} transitions depends strongly on the trap levels and shows several points of vanishing coupling over the range of trap levels with significant population (see \cref{fig:Rabi_freq_low_trap_freq}(b) and \cref{eqn:rabi_freq}). Employing pulses on the \nth{2} order sideband as in the scheme for a single \Mgtf is no longer sufficient. Instead, we employ as many higher order sidebands as necessary. For larger motional states, successively higher order sidebands exhibit a maximum in their coupling rate. Moreover, pulses on higher order sidebands are more efficient since more than one phonon is removed per sideband pulse. For simplicity, we split the total cooling time $T_\mathrm{c}$ equally between all the sideband orders and apply to all of them the same pulse length, thus extending the sequence shown in \cref{fig:seq_sbc_mg25_Scheme1} to higher-order modes. This is a valid approach, since the variation in maximum Rabi frequency across the higher order sidebands is small. Simultaneously to each \gls{RSB} pulse, we apply the quench coupling to further reduce the sensitivity to the optimum pulse length. After each \gls{RSB} pulse, a repumping pulse as described before is applied. The optimal number of applied sideband orders, derived from measurements of the cooling time constant as a function of the maximally applied sideband order,  is shown in \cref{fig:cooling_constant_max_order}{a}. In the optimal sequence, we need to apply sideband pulses up to the \nth{8} order, which confirms the prediction of \cref{fig:Rabi_freq_low_trap_freq}{b}.

More generally, with known Lamb-Dicke parameter and thermal distribution over
motional states after \acrlong{DC}, a corresponding \gls{SBC} strategy can be
used to achieve the most efficient cooling as illustrated in
\cref{fig:cooling_constant_max_order}{b}, where we define the cooling
efficiency as $\beta(n)\cdot \Omega_{n-\beta,n}$.
For a selected upper trap level to be addressed by cooling and a known Lamb-Dicke parameter, the sideband order with
highest cooling efficiency is plotted. 
% \cPiet{Does this plot include the higher
% cooling rate of higher order sidebands, i.e. $\Delta n/T_\pi$? If so, say so
% explicitly.}        
%With a Lamb-Dicke parameter of 0.45 and significant population in trap levels up to $n=120$, \gls{RSB} pulses up to the eighth order are necessary, in agreement with the experimental findings.
This strategy can be combined with multi-mode cooling through interleaved pulse
sequences as shown in \cref{sec:sbc_two_ion_crystal}. 

%####################################################################################
\section{Summary \& discussion}
\label{sec:discussion_sbc}
%####################################################################################
%We find that a pulse length matching the maximum of the RSB excitation rate and an equal amount of time spent on first and second order RSBs provides the fastest cooling (\cref{sec:sbc_of_single_ion}). 
A quasi-continuous sideband cooling scheme was introduced by applying a quench coupling to the excited state in pulsed Raman sideband cooling. For long Raman pulses this scheme allows multiple RSB--spontaneous emission cycles, reminiscent of continuous quench cooling. This feature makes SBC more robust and significantly reduces the optimization required for optimum cooling, since RSB pulses that are longer than the $\pi$-time for a specific initial motional state contribute more efficiently to cooling. 
%The developed scheme is robust against variations of  the Rabi frequency and therefore provides high stability against intensity and pointing fluctuations of the laser, which reduces significantly the optimization required for optimum cooling.
Using this scheme, we performed an optimization of the cooling rate of
single-ion ground state cooling and two-ion sympathetic ground state cooling,
significantly outperforming previous results \cite{hemmerling_single_2011}.
These findings are important to minimize the time required for ground state
cooling, thus reducing overhead in experiments with trapped ions. The results
are applicable to other commonly used hyperfine qubit systems such as 
$^{9}$Be$^+$, $^{111}$Cd$^+$, or $^{171}$Yb$^+$. A variation of the scheme can
be applied to other systems like $^{40}$Ca$^+$, $^{24}$Mg$^+$ in high magnetic
field or optically trapped neutral atoms. 
% \cPiet{What do you have in mind here?}          

We have also demonstrated experimentally that SBC to the ground state is possible outside the Lamb-Dicke regime by employing higher-order RSB transitions, thus confirming theoretical predictions \citep{morigi_ground-state_1997, stevens_simple_1998, morigi_laser_1999}. This regime is particularly relevant for systems where the Doppler cooling linewidth significantly exceeds the motional frequency as is the case for optically trapped neutral atoms, or in situations where the Lamb-Dicke factor is deliberately chosen as large as possible. This is the case for example in experiments which aim to detect small forces through the excitation of the ion's motion from the ground state, such as in photon recoil spectroscopy \cite{wan_precision_2014} and the detection of electric fields \cite{heinzen_quantum-limited_1990, arrington_micro-fabricated_2013}.

The demonstrated motional ground state cooling of a molecular ion, in particular when combined with cooling at low trap frequency, represents an important step towards the implementation of non-destructive state preparation and detection techniques \citep{schmidt_spectroscopy_2006, vogelius_probabilistic_2006, koelemeij_blackbody_2007} based on the detection of small optically-induced and state-selective displacement forces acting on a trapped molecular ion. 

\section*{Acknowledgements}
We acknowledge the support of DFG through QUEST and grant SCHM2678/3-1. Y.W. acknowledges support from IGSM. We thank R. Blatt for generous loan of equipment.

\appendix   

%####################################################################################
\section{Numerical simulation for SBC}
\label{app:numerical_simulation_SBC}
%####################################################################################
\source{The simulation is under
\url{N:/Doktorarbeit_Sync_N10451/Writing/Mg25_CSBC/matlab/simulations}} 
\begin{figure}[b]     
	\centering  
	\includegraphics[width=0.35\textwidth]{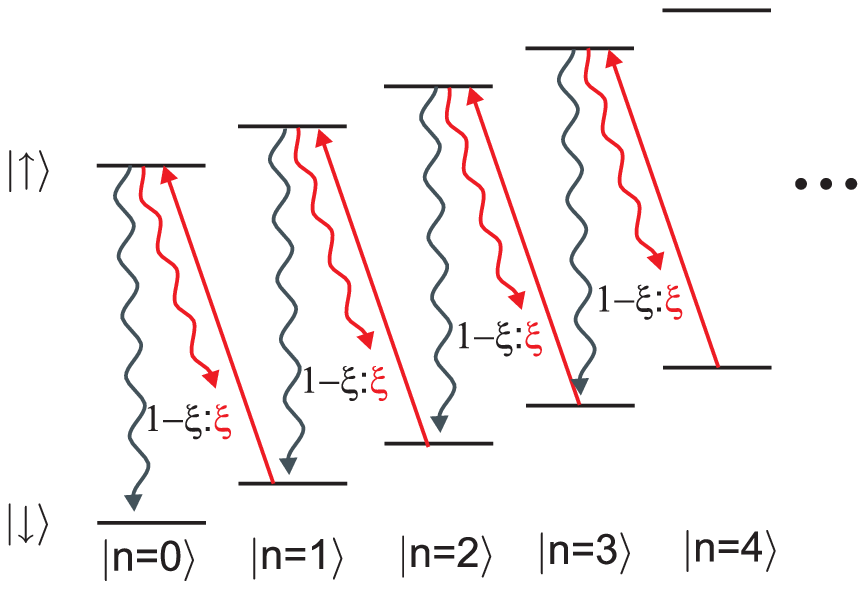}    
	\captionmacro[Relevant levels for numerical simulation of the dynamics during the
	Raman SBC cycles.]{
	The Raman lasers are tuned to be resonant with the $\beta^\mathrm{th}$ order  
	\gls{RSB} ($\beta=1$ in graph).
	Through the dissipative channels either from off-resonant excitation or via
	a repumping process followed by spontaneous emission, the ion
	is reinitialized to the \downket with a decay rate of \SI{0.005}{\per\second}  
	and \SI{0.47}{\per\second}, respectively.     
% 	\cPiet{How are the two different decay
% 	rates implemented in the simulations?}. 
	Spontaneous emission happens on the carrier transition with a probability $1-\xi$ and the remaining small fraction
	$\xi$ happens on the RSB transitions. 
% 	\cPiet{add dots maybe to clarify that more than 4 trap levels are relevant.}
	\source{\url{N:/Doktorarbeit_Sync_N10451/Writing/Mg25_CSBC/Document/media/theory/coupled_bloch_sphere.cdr}}}
	\label{fig:numerical_simulation}
\end{figure}  
% \cPiet{harmonize notation with the main text}
    
The dynamics of the system during quasi-continuous \gls{SBC} is modelled using optical Bloch equations,
 where two electronic states $\upket$ and $\downket$ and 80 trap levels (\cref{fig:numerical_simulation}) are considered.
  The Raman \gls{RSB} pulses are included as a resonant coupling between $\Ket{\downarrow, n}$ 
  and $\Ket{\uparrow, n-\beta}$ with $\beta$ indicating the sideband order used. The spontaneous decay during 
  the Raman cooling cycle via the auxiliary states~\footnote{The spontaneous decay during the Raman cooling cycle is either caused by the off-resonant excitation from the Raman lasers or via the RF coupling and the \dpoh laser.} is implemented
   with an effective decay rate $\Gamma_\mathrm{eff}$ determined experimentally. 
   This approach neglects effects from the actual multi-level electronic structure of
    \Mgtf. Repumping after application of the RSB is implemented through a projection of the
     population onto the electronic ground state \downket. The spatially averaged Lamb-Dicke factor
      for spontaneous emission along the axial direction is
      $\tilde{\eta}=0.134$.
%       \cPiet{How can we claim to be in the Lamb-Dicke regime for spontaneous emission?} 
      Heating from spontaneous emission is taken into account through a
      branching ratio of $1-\xi:\xi$ for emission on the CAR and RSB transition.
%       \cPiet{Please add a few sentences why we can make all these approximations
%       without loosing the main features of the experiment!}     
%It is assumed to include the heating process during repumping and spontaneous emission. 
Here $\xi$ is on the order of $3\cdot\widetilde{\eta}^2\approx 0.05$, where the
factor $3$ considers the multiple scattering events until the ion falls into
the $\downket$ state.
% \cPiet{Where does the factor of 2 come from?}
% %In this way the ion
% % according to the following rule
% % \question{The following equation needs to be changed to diagram.} 
% %\begin{equation}
% %	\Ket{\uparrow, n} \rightarrow (1-\xi)\Ket{\downarrow, n} + \xi\Ket{\downarrow,
% %	n+1}
% %\end{equation}
%in the state $\Ket{\uparrow, n}$ decays into the state $\Ket{\downarrow, n}$ with probability $1-\xi$ and into the state $\Ket{\downarrow, n+1}$ with probability $\xi$. 
Combining all the ingredients above, we end up with the following optical Bloch equations
\begin{align}
	  % Equation 1 
	  \dot{\rho}_\mathrm{\downarrow n, \downarrow n} =&\  i\frac{\Omega_\mathrm{n-\beta,n}}{2}\left[\rho_\mathrm{\downarrow n, \uparrow(n-\beta)} -
	  \rho_\mathrm{\uparrow(n-\beta), \downarrow n}\right]  
	  \notag\\
	  &\ + (1-\xi)\cdot\Gamma_\mathrm{eff}\rho_\mathrm{\uparrow n, \uparrow n} 
	  \notag\\
	  &\ + \xi\cdot\Gamma_\mathrm{eff}\rho_\mathrm{\uparrow(n-1), \uparrow(n-1)} 
	  \label{eqn:rho_gg}
	  \\
	  % Equation 2
	  \dot{\rho}_\mathrm{\downarrow n, \uparrow(n-\beta)} = &\ -\dot{\rho}_\mathrm{\uparrow(n-\beta), \downarrow n} \notag\\
	  = &\ i\frac{\Omega_\mathrm{n-\beta,n}}{2}\left[\rho_\mathrm{\downarrow n, \downarrow n}
	   - \rho_\mathrm{\uparrow(n-\beta), \uparrow(n-\beta)}\right] 
	   \notag\\
	  %- i\delta \rho_\mathrm{\downarrow n, \uparrow(n-\beta)} &
	  &\ - \frac{\Gamma_\mathrm{eff}}{2}\rho_\mathrm{\downarrow n, \uparrow(n-\beta)} \\
	  % Equatin 3
%	  \dot{\rho}_\mathrm{\uparrow(n-\beta), \downarrow n} = &\ i\frac{\Omega_\mathrm{n-\beta,n}}{2}\left[\rho_\mathrm{\uparrow(n-\beta), \uparrow(n-\beta)} - \rho_\mathrm{\downarrow n, \downarrow n}\right] \notag\\
	  % + i\delta \rho_\mathrm{\uparrow(n-\beta), \downarrow n} &
%	  &\ - \frac{\Gamma}{2}\rho_\mathrm{\uparrow(n-\beta), \downarrow n} \\
	  % Equation 4
	  \dot{\rho}_\mathrm{\uparrow(n-\beta), \uparrow(n-\beta)} = &\
	  i\frac{\Omega_\mathrm{n-\beta,n}}{2}\left[\rho_\mathrm{\uparrow(n-\beta), \downarrow n} -
	  \rho_\mathrm{gn, \uparrow(n-\beta)}\right] \notag\\
	  &\ - \Gamma_\mathrm{eff}\rho_\mathrm{\uparrow(n-\beta),
	  \uparrow(n-\beta)}
	  \label{eqn:rho_ee}
\end{align}
with the density matrix elements defined in the usual way, e.g. 
\begin{align}
%	 \rho_\mat\downarrow nm{\downarrow n,\downarrow n} =&\ \Bra{g, n} \rho  \Ket{g, n} \\\
	 \rho_\mathrm{\downarrow n, \uparrow(n-\beta)}
	 = &\
	 \Bra{\downarrow, n} \rho  \Ket{ \uparrow, (n-\beta)}
%	 \rho_\mathrm{\uparrow(n-\beta), \downarrow n} =& \ \Bra{e, (n-\beta)} \rho  \Ket{g, n} \\  														
%	 \rho_\mathrm{\uparrow(n-\beta),\uparrow(n-\beta)} =&\ \Bra{e, (n-\beta)} \rho  \Ket{e, (n-\beta)}
\end{align}
and $\Omega_\mathrm{n,n-\beta}$ as the effective Rabi frequency coupling the two states $\Ket{\downarrow, n}$ and $\Ket{\uparrow, n-\beta}$. 
This system of differential equations is solved numerically to produce the
evolution during a single Raman cooling pulse. 
% \cPiet{Add a sentence how the
% repump pulse after the RSB pulse is implemented.}      

For every set of parameters ($T_\mathrm{c}$, $\alpha$, $t_\mathrm{R1}$ and $t_\mathrm{R2}$), the evolution during the pulse sequence as shown in \cref{fig:seq_sbc_mg25_Scheme1} is computed by repeatedly solving the equations above. The actual evolution time for the atomic system during each pulse is reduced by \SI{1}{\mus} from the chosen pulse lengths to include realistic experimental pulse area reduction. The final population in the motional ground state is considered as the signal detected by the \gls{STIRAP} pulse and corrected by an amplitude $a = 0.7$ and an offset $b = 0.23$ reduction due to experimental imperfections according to 
\begin{equation}
y = a\times(y_\mathrm{theory}-b).
\end{equation}
A scan over the \gls{SBC} cooling time $T_\mathrm{c}$ reproduces the experimental data, which are processed in a similar way as described in \cref{sec:sgl_Mg_scheme}.

%\bibliography{Script_CSBC_Mg25} 
\bibliography{SBC}               
 
\end{document}